\documentclass{aa}
\usepackage{graphicx}
\usepackage{txfonts}
\newcommand{\beq}{\begin{equation}}
\newcommand{\eeq}{\end{equation}}
\newcommand{\beqn}{\begin{eqnarray}}
\newcommand{\eeqn}{\end{eqnarray}}
\newcommand{\gppr}{\stackrel{>}{\scriptstyle \sim}}
\newcommand{\lppr}{\stackrel{<}{\scriptstyle \sim}}
\begin{document}
   \title{On particle acceleration and very high energy
   $\gamma$-ray emission in Crab-like pulsars}

   \author{Osmanov Zaza
          \inst{1}\fnmsep\thanks{z.osmanov@astro-ge.org}
 \and
          Rieger Frank M. \inst{2,3}
          }

   \offprints{Zaza Osmanov}

   \institute{E. Kharadze Georgian National Astrophysical
   Observatory, Ilia Chavchavadze State University,
              Kazbegi str. 2a, 0106 Tbilisi, Georgia;
              \email{z.osmanov@astro-ge.org}
         \and Max-Planck-Institut f\"ur Kernphysik,
              Saupfercheckweg 1, 69117 Heidelberg, Germany;
              \email{frank.rieger@mpi-hd.mpg.de}
         \and European Associated Laboratory for Gamma-Ray Astronomy,
              jointly supported by CNRS and MPG
              }



  \abstract
  {The origin of very energetic charged particles and the production of very high-energy (VHE)
  gamma-ray  emission remains still a challenging issue in modern pulsar physics.}
   {By applying a toy model, we explore the acceleration of co-rotating charged particles
     close to the light surface in a plasma-rich pulsar magnetosphere and study their interactions
     with magnetic and photon fields under conditions appropriate for Crab-type pulsars.}
   {Centrifugal acceleration of particles in a monopol-like magnetic field geometry is analyzed and
     the efficiency constraints, imposed by corotation, inverse Compton interactions and curvature
     radiation reaction are determined. We derive expressions for the maximum particle energy and
     provide estimates for the corresponding high-energy curvature and inverse Compton power outputs.}
   {It is shown that for Crab-like pulsars, electron Lorentz factor up to $\gamma \sim 10^7$ can be
    achieved, allowing inverse Compton (Klein-Nishina) up-scattering of thermal photons to TeV
    energies with a maximum luminosity output of $\sim10^{31}$ erg/s. Curvature radiation, on the
    other hand, will result in a strong GeV emission output of up to $\sim(10^{34}-10^{35})$ erg/s,
    quasi-exponentially decreasing towards higher energies for photon energies below $\sim 50$
    GeV.}
   {Accordingly to the results presented only young pulsars are expected to be sites of detectable
    VHE $\gamma$-ray emission.}

   \keywords{Pulsars -- Acceleration of particles -- emission processes}

   \maketitle
%


\section{Introduction}
One of the fundamental problems in pulsar physics is related to the
origin of the observed non-thermal emission. While it seems evident
that efficient particle acceleration and emission processes must be
operating in a pulsar's rotating magnetosphere, current theoretical
approaches differ widely in their assumptions about the localization
of the relevant zones: According to standard polar cap models, for
example, charged particles are uprooted from the neutron star's
surface by strong electrostatic fields (\cite{rud}). Still close to
the star, these particles are then assumed to be efficiently
accelerated along open field lines in parallel electric fields
induced by space-charge-limited flow (\cite{michel}), field-line
curvature (\cite{arons}) and/or inertial frame dragging effects
(\cite{musl}). In most cases, the parallel electric field component
is shorted out at some altitude by the onset of electron-positron
pair cascades in strong magnetic fields (one-photon pair
production), either initiated by curvature(\cite{dau}) or Inverse
Compton radiation (\cite{der}). Outer gap models, on the other hand,
assume that primary particles are efficiently accelerated in vacuum
gaps in the outer magnetosphere, inducing pair cascades through
$\gamma\gamma$-pair production (\cite{cheng, chiang, hirotani}). In
all these approaches the maximum attainable particle energy is
either limited by the gap size or radiation reaction. In
conventional polar cap models, for example, a critical issue has
always been the question whether particles can indeed gain enough
energy inside the gap to  account for the observed non-thermal
radiation from $\gamma$-ray pulsars. Several scenarios have been
proposed to enlarge the gap zone and consequently increase the
corresponding energy output (e.g., \cite{usov, arons1, musl}), yet
$\gamma$-ray emission from Crab-like pulsars still proves
challenging to account for. The efficiency of particle acceleration
along magnetic field lines has also been studied more recently based
on numerical solutions of the structure of a stationary,
axisymmetric and force-free magnetosphere of an aligned pulsar
(\cite{contop}). According to the results obtained, the relativistic
magnetospheric outflow is not accelerated efficiently enough to
account for the production of high energy gamma-rays. In some
respects, this result may not come unexpected as the magnetic field
configuration is restricted to be force-free, thus preventing
efficient acceleration. In a rather different approach, \cite{besk}
analyzed the acceleration of a (stationary) two-component,
electron-positron outflow in a monopole magnetic field configuration
for high (Michel) magnetization parameters $\sigma\equiv B^2/8\pi n
m_e c^2$, indicating the ability of the field to sling particles to
high velocities. Here, $n_e$ is the electron density, $B$ the
induction of the magnetic field, $c$ is the speed of light and $e$
and $m_e$ are electron's charge and the rest mass respectively.
Considering plasma dynamics close to the force-free regime
(first-order correction), they showed that for small longitudinal
currents very high Lorentz factors can be achieved, with almost all
of the electromagnetic energy being converted into the kinetic
energy of particles ($\gamma_{\rm max} \sim \sigma$) in a thin layer
close to the light cylinder surface.

In the present contribution we consider another acceleration
mechanism that may help to overcome the energy problem arising in
some polar cap-type models. To this end, we explore the acceleration
of co-rotating charged particles in an idealized monopole-like
magnetic field region close to the light surface where the parallel
electric field component is effectively screened  out, but where
inertial (centrifugal) effects become important in describing the
plasma dynamics. This follows earlier suggestions by
Gold~(\cite{gold1,gold2}) about efficient particle acceleration
close the light cylinder in a co-rotating neutron star magnetosphere
(see also \cite{rud72}). A detailed analysis of centrifugal
acceleration along rotating straight field lines in the test
particle limit has been presented by Machabeli \&
Rogava~(\cite{mr94}), showing that due to the relativistic mass
increment the radial acceleration of a particle changes sign,
similar to results obtained for particle motion close to a
Schwarzschild black hole (Abramowicz \& Prasanna \cite{abr}). Based
on this, the plasma motion in pulsar magnetospheres has been
analyzed and equations describing the behavior of a co-rotating
plasma stream have been derived (e.g., \cite{che96, mach05}). More
recently, the generalization to curved field lines (e.g., Archimedes
spiral, where a particle may asymptotically reach the force-free
regime) has been examined and the consequences of radiation reaction
analyzed (\cite{r03, dalak}). Independently, application{\bf s} of
centrifugal particle acceleration to milli-second pulsars were
considered and curvature radiation effects discussed in Gangadhara
(\cite{gang96}) (see also \cite {thom07} for a recent
generalization). In a wider context, the efficiency of centrifugal
particle acceleration was studied for Active Galactic Nuclei (AGN)
(\cite{gl97, rm00, osm7}), based on scenarios where AGN jets
originate as centrifugally-driven outflows (\cite{bp82}).

In the present paper, we analyze the efficiency of centrifugal acceleration
for Crab-like pulsars, taking constraints imposed by co-rotation, inverse
Compton interactions and curvature radiation into account. The paper is
arranged as follows: In Sec. 2 the radial particle motion due to
centrifugal acceleration effects is described and co-rotation constraints
discussed. In Sec. 3 we examine possible radiative feedbacks on
the process of acceleration for typical millisecond pulsars, considering
some major limiting processes: inverse Compton scattering, curvature
radiation and pair creation. In Sec. 4 the relevance of our results is
shortly discussed in the context of recent observational evidence.

\section{Centrifugal acceleration of particles}

\subsection{A simplified approach}
We consider an idealized, single-particle approach, where in the
local frame of reference, each particle is only subject to the
action of the centrifugal force and gains energy while moving
outward along the magnetic field. The field configuration is
supposed to be almost straight, at least inside the co-rotation
zone. This seems a reasonable assumption to make as efficient
particle acceleration is expected to take place on characteristic
length scales much smaller than the light cylinder radius. When
applied to open field lines, our results are thus not expected to be
very sensitive to further magnetospheric details. If $\alpha$
denotes the angle between the magnetic field $\vec{B}$ and angular
velocity $\vec{\Omega}$, then the effective field line rotation is
$\Omega_e=\Omega \sin \alpha$ and the light surface is at a distance
$r_{\rm L}=r_{\rm L,0}/\sin\alpha$, where $r_{\rm L,0}=c P/2\pi$ is
the light cylinder radius and $P=2\pi/\Omega$ is the pulsar period.
The dynamics of a co-rotating particle can then be described by the
following equation (Machabeli \& Rogava \cite{mr94}, \cite{rm00}):
\begin{equation}
\label{eul_0}
\frac{d^2r}{dt^2}=\frac{\Omega_{e}^2r}{1-\Omega_{e}^2r^2}\left[1-\Omega_{e}^2r^2-2
\left(\frac{dr}{dt}\right)^2\right],
\end{equation}
and its Lorentz factor can be expressed as
\begin{equation}\label{gamma}
\gamma(r) = \frac{1}{\sqrt{\tilde{m}}~(1-r^2/r_{\rm L}^2)}\,,
\end{equation}
where $\tilde{m} \simeq 1/\gamma_0^2$ is essentially determined by the
initial conditions. Hence, for a secondary pair plasma produced close to
the neutron star with Lorentz factor $\gamma_0 \sim (10^3-10^4)$, for
example, efficient centrifugal acceleration to high energies acceleration
can take place close to the light surface, i.e., in a layer $\Delta r/r_{\rm L} \sim
\gamma_0/\gamma$. Using Eq.~(\ref{gamma}), the characteristic timescale
for centrifugal acceleration can be approximated by
\beq\label{accel_time}
t_{\rm acc}\equiv \frac{\gamma}{d\gamma/dt} \simeq \frac{P}{4 \pi
\sin\alpha\,\tilde{m}^{1/4}\gamma^{1/2}}\,.
\eeq
Obviously, for a particle approaching the light surface, the acceleration
timescale decreases with $\gamma$ and the Lorentz factor can increase
dramatically unless co-rotation can no longer be maintained or radiation
reaction becomes important.

\subsection{Co-rotation constraints}
Suppose that the co-rotation zone extends outwards from the neutron
star up to the vicinity of the the light surface (Gold 1968, 1969). Because
of strong synchrotron losses, electrons will quickly lose their relativistic
perpendicular energy, i.e. on a timescale which for most pitch angles $\psi$
is much smaller than the transit time$\tau_t=\gamma^2 \sin^2\psi~ t_s$ to
their ground Landau state, so that they may be approximately described as
moving one-dimensionally along the field lines. Yet, even if one neglects
radiation reaction (e.g., curvature losses, see below) co-rotation will only
be possible as long as the kinetic energy density of the electrons $\gamma
n m_e c^2$ does not exceed the energy density in the field $B^2/8\pi$
(Alfv\'en corotation condition). For a number density $n=M\,n_{\rm GJ}$,
where $M$ denotes the multiplicity (number of secondaries to number of
primaries) and $n_{\rm GJ}= \Omega B \cos\alpha/(2\pi e c) = 0.07~B \cos
\alpha/P$ [particles~cm$^{-3}$] the classical Goldreich-Julian number
density close to the star, the co-rotation condition implies an upper
limit for achievable electron Lorentz factors of
\beq\label{corotation} \gamma_{\rm max, e}^{\rm cor} \simeq 4 \times
10^5 \frac{B P}{M \cos\alpha}\,.
 \eeq
 For a Crab-type pulsar with $P\sim 0.033$ s and $B\simeq 10^6 \sin^3
 \alpha$ G (at the light surface $r_{\rm L}$, assuming an internal
 dipolar field structure), for example, this results in $\gamma_{\rm max,
 e}^{\rm cor}  \simeq 2 \times 10^{10} \tan\alpha\sin^2
 \alpha/M$, while for a 1s pulsar ($B \sim 10 \sin^3 \alpha$ G at $r_{\rm L}$)
 the upper limit would be of the order of $\gamma_{\rm max, e}^{\rm cor}
 \simeq 4\times 10^7 \tan\alpha \sin^2\alpha/M.$ These values suggest
 that for milli-second pulsars co-rotation may indeed last up to the very vicinity
of the light cylinder. Note that close to the light surface, the number density
required to screen out a potential parallel electric field component is much
larger than $n_{\rm GJ}$ employed above, and in fact given by (Goldreich
\& Julian~1969)
\beq\label{GJ} n_{\rm GJ,L}^L = n_{\rm GJ}
                           \frac{1}{(1-\Omega_e^2 r^2/c^2)}.
\eeq
The onset of a pair production front in polar cap models, close to the neutron
star, is usually expected to result in a multiplicity $M =n/n_{\rm GJ}\sim 10^2-
10^5$ and a mean Lorentz factors of the secondaries of $\gamma_0 \sim
10^3-10^4$ (e.g., \cite{dau1}; \cite{mel98}; Baring 2004).
On the other hand, using Eq.~(\ref{gamma}), Eq.~(\ref{GJ}) can be
expressed as $n_{\rm GJ,L} \simeq n_{\rm GJ}\,\gamma/\gamma_0$. Hence,
in order to achieve electron Lorentz factors of, e.g., $\gamma \sim 10^7$ via
centrifugal acceleration, one requires $M \sim \gamma/\gamma_0 \sim (10^3
-10^4)$, which seems well possible given the parameter range above.
Equation~(\ref{corotation}) then suggests conversely, that for Crab-type pulsars,
Lorentz factors up to $\gamma_{\rm max,e} \sim 10^7$ can indeed be achieved
if radiation reaction is negligible. Acceleration will then occur in a narrow
region very close to the light surface with $\Delta r/r_{\rm L} \sim 1/M$. This
seems reminiscent of earlier results about particle acceleration in the presence
of small longitudinal currents (Beskin et al. 1983; Beskin \& Rafikov 2000).

\section{Emission constraints}
In realistic astrophysical environments, radiation reaction will impose
additional constraints on the efficiency of any particle acceleration process.
For pulsars, important limitations could arise through inverse Compton
scattering with ambient soft photons field or synchro-curvature losses
along curved particle trajectories. Pair production (i.e., one-photon or
photon-photon) on the other hand, could possibly lead to a suppression
of detectable high energy $\gamma$-rays.

\subsection{Inverse Compton interactions}
Thermal radiation as well as synchrotron radiation by secondary electrons
could in principle lead to a non-negligible target photon field for Inverse
Compton (IC) interactions and thereby limit achievable electron energies.

\subsubsection{Inverse Compton with thermal photons}
It has often been assumed that IC interactions with thermal photons
from the neutron star surface are generally negligible far away from
the surface because (i) the photon density decreases with distance
$r$ and (ii) charges and photon are traveling in almost the same
direction, so that (anisotropic) inverse Compton losses become
exceedingly small (e.g., \cite{mor}). While the first consideration
is certainly true, the latter may not necessarily be the case. In
fact, if electrons are co-rotating with the plasma, their main
velocity component close to the light cylinder is expected to be in
the azimuthal direction, implying a preferred interaction angle of
almost 90 degree, so that IC interactions with thermal photon field
may possibly become relevant for milli-second pulsars. Although
pulsars are born at very high temperatures $T\sim 10^{11}$ K, their
surface temperatures quickly cool down to $\sim 5 \times 10^6$ K by
various neutrino emission processes and thermal emission of photons
(e.g., \cite{tsu}; \cite{yakpet}). Standard (modified UCRA, plasma
neutrino and photon cooling) models predict a surface temperature
above $10^6$ K for pulsars with ages $\tau=P/2\dot{P} \lppr 10^{4}$
yr (neutrino cooling stage), and below $10^5$ K for pulsars
exceeding $10^{6.7}$ yr (photon cooling stage). Using the standard
cooling curve, one can employ an approximate phenomenological
description for the temperature-age dependence given by
(\cite{zhang})
\beqn T(\tau)  &\simeq& 5.9 \times
10^5~\mathrm{K} \left(\frac{10^6~\mathrm{yr}}{\tau}\right)^{0.1}
\quad \mathrm{for}\; \tau \leq 10^{5.2} \mathrm{yr}\,,\\
T(\tau)  &\simeq& 2.8 \times 10^5~\mathrm{K}
\left(\frac{10^6~\mathrm{yr}}{\tau}\right)^{0.5} \quad
\mathrm{for}\; \tau  > 10^{5.2} \mathrm{yr}\,.
\eeqn
Hence, for a Crab-type pulsar ($\tau \simeq 10^3$ yr, $r_s/r_{\rm L,0} \simeq
10^{-2}$) (where $r_s$ is the stellar radius) the surface temperature may be
approximated by $T \simeq 1.2 \times 10^6$ K. The Planck function then
peaks at around $\nu_{\rm max} =2.8~k T/h \simeq 7 \times 10^{16}$ Hz,
corresponding to a photon energy $\epsilon_{\rm ph} \simeq 0.3$ keV. IC
scattering thus mainly occurs in the extreme Klein-Nishina regime. As a first
order approximation for the single particle (non-resonant) Klein-Nishina
Compton power, one can employ the expression derived by \cite{blum}
assuming a (quasi-isotropic) black -body photon distribution, i.e.,
 \beq
P_{\rm c, KN} \simeq \frac{\sigma_T (m_e c k T)^2}{16 \hbar^3}
\left(\ln \frac{4 \gamma k T}{m_e c^2} - 1.981\right)
                         \left(\frac{r_s}{r}\right)^2\,,
\eeq
where $\gamma$ is the electron Lorentz factor. This implies a characteristic
IC cooling timescale close to of $r_{\rm L}$ of
\beq
t_{\rm IC} = \frac{\gamma m_e c^2}{P_{\rm c, KN}} \propto \gamma\,,
\eeq
which to first order is proportional to $\gamma$. Acceleration, on the other
hand, occurs on a timescale $t_{\rm acc} \propto 1/\gamma^{1/2}$, so that
for electron Lorentz factors $\gamma \gppr \gamma_0$, IC cooling will not
impose any constraints on achievable particle energies. This suggests
that for Crab-like pulsars electron Lorentz factors are essentially limited by
co-rotation and not by IC radiation reaction (cf. eq.~\ref{corotation}). If so,
then detectable IC emission at $\gamma m_e c^2\sim 5~(\gamma/10^7)$ TeV
may occur.
We can roughly estimate the possible TeV luminosity by multiplying the
corresponding particle number with the single IC power $P_{\rm c, KN}$,
i.e.,  $L_{\rm IC}^{\rm TeV} \sim n_{\rm GJ} M  \Delta V P_{\rm c, KN}$, which
gives
\begin{equation}
L_{\rm IC}^{\rm TeV}\sim 10^{31}
\left(\frac{B}{10^6~\mathrm{G}}\right) \left(\frac{T}{1.2 \times
10^6~\mathrm{K}}\right)^2 \left(\frac{\chi}{\sin\alpha}\right)\;\;
\mathrm{erg/s},
\end{equation}
where  $\Delta V = 4 \pi \chi r_{\rm L}^2 d \sim 2 \pi \chi r_{\rm L}^3/M$, with
$\chi \leq 1$ denoting the deviation from isotropy and $d \sim (\gamma_0/ 2
\gamma)~r_{\rm L}$ the thickness of the layer close to $r_{\rm L}$, in
which the highest particle energies are achieved.  The (pulsed, non-steady)
TeV luminosity could thus be as  high as $\sim10^{31}$ erg/s, consistent
with existing upper limits derived by current ground-based $\gamma$-ray
instruments (IACT) (e.g., \cite{less}; \cite{ahar}; \cite{albert}), yet possibly
accessible to the next-generation CTA-type instruments.

\subsubsection{Inverse Compton with infrared photons}
Secondary synchrotron emission could possibly lead to a non-negligible
photon field in the infrared-optical regime where IC interactions may occur
in the Thomson regime. For the Crab pulsar, the (isotropic, phase-averaged)
near infrared-optical luminosity is of order $L_{\rm o} \sim 10^{33}$ erg/s,
turning significantly downward for lower frequencies  (e.g., \cite{middl};
\cite{eiken}; \cite{soller}). This suggests a photon energy density close to the
light surface of order $u_{\rm ph} \sim L_{\rm o}/(4\pi r_{\rm L} c) \sim 10^5
\sin^2\alpha$ erg/cm$^3$, comparable to the thermal one. Approximating
the single particle (non-resonant, quasi-isotropic) Compton power by $P_c
\sim \sigma_T c  \gamma^2 u_{\rm ph}/(1+x)$, where $x=\gamma
\epsilon_{\rm ph}/m_e c^2$, the characteristic IC cooling timescale
close to $r_{\rm L}$ then becomes
\beq t_{\rm IC} \simeq \frac{\gamma m_e c^2}{P_c} \simeq
                          \frac{4.1 \times 10^7\,\mathrm{sec}}{\gamma~u_{\rm ph,L}}
                         (1 + 2 \gamma/10^{6})\,.
\eeq Comparing acceleration, occurring on $t_{\rm acc}$ (Eq.~\ref{accel_time}),
with IC cooling, occurring on $t_{\rm IC}$, implies
\beq\label{IC_limit} \frac{\gamma}{(1+2 \gamma/10^6)^2}
\lppr \frac{2 \times 10^{10}} {\gamma_0 \sin^2\alpha}
\eeq and verifies that IC interactions with the infrared-optical photon field
will not impose a severe constraint on the maximum achievable Lorentz
factor. IC up-scattering by electrons with $\gamma \sim 10^7$ would again
produce emission at around $5$ TeV. If the observed NIR-optical emission
would originate from regions close to the pulsar, IC scattering close to
$r_{\rm L}$ might be reasonably approximated by the quasi-isotropic
expression $P_c$. This would result in a possible TeV output $L_{\rm IC}
\sim n_{\rm GJ} M  \Delta V P_{\rm c} \sim 10^{35} {\bf \chi} \sin \alpha
\cos\alpha$ erg/s, exceeding the existing upper limits noted above.
On the other hand, if the NIR-optical emission is produced close to
$r_{\rm L}$ as proposed in some models (e.g., \cite{pacini}; \cite{crus}),
anisotropic inverse Compton scattering may severely reduce this power
output (by up to a factor $\psi^4$ where $\psi \ll 1$ is the pitch angle, cf.
\cite{mor}), suggesting a TeV contribution well below the one produced
by thermal IC. The empirical scaling of the near infrared-optical (and
$\gamma$-ray) flux with the magnetic field at the light cylinder seems
in fact to reinforce a scenario where the emitting region is located close
to the light surface (\cite{shear}).

\subsection{Curvature radiation}
Approaching the light surface, field line bending may no longer be
negligible so that a particle may efficiently lose energy due to curvature
radiation. In analogy to synchrotron radiation, curvature radiation can
be described as emission from relativistic charged particles moving
around the arc of a circle, chosen such that the actual acceleration
corresponds to the centripetal one (e.g., \cite{ochusov}). The critical
frequency where most of the radiation is emitted is given by
\beq\label{curv_freq} \nu_c \simeq
\frac{3 c}{4\pi R_c} \gamma^3\,,
\eeq which for, e.g., $R_c \sim r_{\rm L} \sim 10^8$ cm (cf. Gold
\cite{gold1,gold2}) yields
$\nu_c \simeq 5 \times
10^{22}~(\gamma/10^7)^3$ Hz or a curvature photon energy of about
$0.2~(\gamma/10^7)^3$ GeV. The energy loss rate or total power
radiated away by a single particle is
\beq\label{curv_power}
  P_c = \frac{2}{3} \frac{e^2 c}{R_c^2} \gamma^4\,.
\eeq The characteristic cooling timescale $t_c = \gamma m_0 c^2/P_c$
thus becomes
\beq
  t_c \simeq 180 R_c^2 \left(\frac{m_0}{m_e}\right) \frac{1}{\gamma^3}\,.
\eeq To find the maximum electron Lorentz factor attainable in the
presence of curvature radiation, we can again balance $t_{\rm acc}$
(Eq.~\ref{accel_time}) with $t_{c}$ to obtain
\beq\label{curvgama}
 \gamma_{\rm max} \simeq 1.2 \times 10^9 P^{2/5}
 \left(\frac{R_c}{r_{\rm L}}\right)^{4/5} \frac{1}{\gamma_0^{1/5}}\sin^{2/5}
 \alpha\,,
\eeq indicating that for a Crab-type pulsar ($P=0.033$ s, $R_c \sim
r_{\rm L}$) with,.e.g., $\gamma_0 \sim 10^4$ curvature radiation constrains
achievable electron Lorentz factors to $\gamma \lppr 4.8 \times 10^7
\sin^{2/5} \alpha$. This confirms that for Crab-type parameters,
electron Lorentz factors up to $\sim 10^7$ might in fact be obtained.
We can again estimate the possible curvature output $L_c^{\rm
GeV}$ at GeV energies, by multiplying the particle number with the
single particle curvature power $P_{\rm c}$ to obtain
\beqn L_c ^{\rm GeV} \sim
n_{\rm GJ} M  \Delta V P_{\rm c} \sim 10^{34}
\left(\frac{B}{10^6~\mathrm{G}}\right)
\left(\frac{r_{\rm L,0}}{R_c}\right)^2 \left(\frac{\gamma}{10^7}\right)^4
\frac{\chi\,\cot\alpha}{\sin^2\alpha}\;\; \mathrm{erg/s}  \nonumber \\
\eeqn using $\Delta V$ (with $\chi \leq 1$) as defined above. Note
that curvature radiation in principle results in an additional IC
(Klein-Nishina) photon target field close to $r_{\rm L}$ with energy
density $u_{\rm ph, c} \sim L_c^{\rm GeV}/(4\pi r_{\rm L}^2 c)$. For
Crab-type values, the resultant energy density would be a factor of
a few higher than the thermal one. Yet, due to both, the further reduced
Klein-Nishina cross-section and anisotropic scattering conditions,
this curvature photon field is not expected to significantly modify our
IC considerations above.

\subsection{Pair creation and $\gamma\gamma$-absorption}
In the magnetosphere of a pulsar, pair creation via magnetic photon
absorption ($\gamma + \vec{B} \rightarrow e^{+} + e^- + \vec{B}$)
becomes kinematically possible as the magnetic field can absorb
momentum and thereby ensure momentum conservation. For a
photon, traveling with pitch angle $\psi$ to the local magnetic field,
the threshold conditions for this being possible is $\epsilon_{\rm ph}
\sin\psi\geq 2 m_e c^2$. The absorption coefficient for this process
is $\kappa \simeq 1.4 \times 10^{-13} (B \sin\psi/B_c)~T(\lambda)$
[1/cm] where $B_c = m_e c^3/(e \hbar) \simeq 4.4 \times 10^{13}$ G,
$\lambda = 1.5~(B \sin\psi/B_c) (h\nu/m_e c^2)$ and $T(\lambda)$ is
the Erber function (\cite{erber}; \cite{tsai}). The function $T(\lambda)$
is very sensitive to $\lambda$. While $\lambda$ might be high near
to the stellar surface, only moderate values are expected close to the
light surface. For small $\lambda \ll 1$ one has $T(\lambda) \simeq
0.46 \exp(-4/\lambda)$ (cf. also \cite{dau1}), while for large $\lambda
\gg 1$ one finds $T(\lambda)\simeq 0.9~\lambda^{-1/3}$. The location
of the absorbing surface may thus be approximated by $\kappa r
\simeq 1$, which gives
\beq
 \lambda^{-1} \simeq -0.25 \ln[6.4 \times 10^{-14} (B/B_c)~r]\,.
\eeq For a Crab-type pulsar ($B\sim 10^6 \sin^3\alpha$ G close to
$r_{\rm L}$) one finds $\lambda^{-1} \simeq 7.6$, so that a photon with
energy above the cut-off $\epsilon_{\rm ph} = h\nu \sim
2/(\sin^3\alpha~\sin\psi)$ TeV will undergo magnetic absorption. As
the pitch angle is usually very small ($\psi \lppr 10^{-2}$), we do
not expect this of significance for Crab-type pulsars at photon
energies below $50$ TeV.

Apart from one-photon pair production, energetic photons may also
undergo photon-photon interactions ($\gamma + \gamma_s \rightarrow
e^{+} + e^-$) with background soft photons of energy $\epsilon_s$
(e.g., \cite{chiang}).
Let us thus consider the following cases:\\
(1) In the case of TeV photons, the threshold condition requires the
presence of soft photons with energies $\epsilon_s = 2 (m_e
c^2)^2/\epsilon = 0.5~(1~\mathrm{TeV}/\epsilon)$ eV or larger. The
cross-section for $\gamma\gamma$-pair production has a sharp
maximum of $\sim 0.2 \sigma_T$ at $2 \epsilon_s$, so that the optical
depth can be approximated by $\tau(\epsilon) \sim 0.1 \sigma_T L(2
\epsilon_s)~ \Delta r/(4 \pi r^2 c \epsilon_s)$ where $L(2\epsilon_s)$
is the corresponding luminosity at which the peak occurs and $\Delta r
\simeq (r-r_{\rm L})$ the path length. For a characteristic (observed
pulsed Crab) photon field of $L(0.1\mathrm{eV}) \ll L(1 \mathrm{eV})
\lppr (1-3) \times 10^{33}$ erg/s (e.g., \cite{eiken}; \cite{soller}) this
would result in
\beq\label{absorption} \tau(\epsilon) \simeq 1
\left(\frac{L(1~\mathrm{eV} [1~\mathrm{TeV}/ \epsilon])}{3 \times
10^{33} \mathrm{erg/s}}\right)
                                           \left(\frac{\epsilon}{1~\mathrm{TeV}}\right)\,
                                           \sin\alpha\,.
\eeq noting that $(r_{\rm L}/r) (1-r_{\rm L}/r) \leq 1/4$ for $r
\geq r_{\rm L}$. Equation~(\ref{absorption}) thus suggests that a
substantial fraction of TeV photons (if not all) may in fact be able
to escape absorption.\\
(2) The situation could be somewhat different for the GeV curvature
photons. In this case $\gamma\gamma$-absorption requires the
presence of soft photons in the X-ray regime with $\epsilon_s =
0.5~(1~\mathrm{GeV}/\epsilon)$ keV or larger. In the case of, e.g.,
the Crab pulsar, the X-ray luminosity is of order $L_x \sim 10^{36}$
erg/s (\cite{kuiper}; \cite{possenti}; \cite{mas06}), so that the optical
depth becomes
\beq \tau(\epsilon) \simeq 0.34
\left(\frac{L(1~\mathrm{keV} [1~\mathrm{GeV}/ \epsilon])}{10^{36}
~\mathrm{erg/s}}\right)
\left(\frac{\epsilon}{1~\mathrm{GeV}}\right)\,\sin\alpha\,.
\eeq To first order, the soft X-ray flux of the Crab follows $L(E) \simeq
10^{36}~(10~\mathrm{keV}/E)^{-0.4}$ erg/s (e.g., \cite{mas00};
\cite{kuiper}). Hence, for photons of energy $\epsilon \sim 20$ GeV
or $50$ GeV, the relevant target luminosity would be $L(2\epsilon_s)
\simeq 1.2 \times 10^{35}$ and $\simeq 8.3 \times 10^{34}$ erg/s
respectively, implying an optical depth $\tau \simeq 0.8 \sin \alpha$
and $\tau \simeq 1.4 \sin \alpha$. As curvature radiation typically
results in a quasi-exponential GeV tail, no significant super-exponential
suppression is expected to occur at energies below several tens
of GeV, which seems consistent with recent results based on the
detection of pulsed emission above 25 GeV from the Crab (\cite{aliu}).

\begin{figure}
  \resizebox{\hsize}{!}{\includegraphics[angle=0]{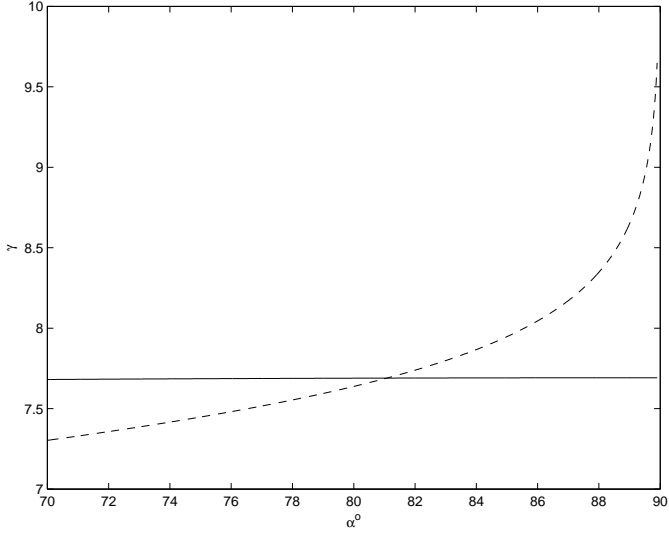}}
  \caption{Maximum Lorentz factors versus the inclination angle:
  $\gamma^{\rm cor}_{\rm max}$ (from co-rotation) and $\gamma^{\rm cr}_{\rm max}$
  (curvature radiation) are represented by the dashed and the solid line respectively.
  The set of parameters emplyed is: $M = 5 \times 10^3$, $P = 0.033$s,
  $\gamma_0 = 10^4$.}\label{fig}
\end{figure}

\section{Conclusions}
Based on our idealized model analysis above, the efficiency of particle
acceleration close to the light surface is essentially limited by curvature
radiation reaction or the Alfv\'en corotation condition.\\
We have studied the implications for conditions applicable to
Crab-type pulsars, assuming a plasma-rich environment with pair
density exceeding the primary Goldreich-Julian (close to the star)
one by $M \sim 10^3$. In this case, electron Lorentz factors up to
$\gamma \sim 2 \times 10^7$ appear possible. Synchro-curvature
radiation could then lead to a relatively strong (averaged) power
output of $\sim (10^{34}-10^{35})$ erg/s at $\sim 2$ GeV that would
be consistent with, e.g., EGRET observations of the Crab
(\cite{kuiper}). The emissivity of curvature radiation can be
described by \beq\label{p}
p(\nu)=\frac{1}{2}\sqrt{\frac{3}{\pi}}\frac{e^2}{R_c}\gamma
\frac{\nu}{\nu_c}\int_{\nu/\nu_c}^{\infty}\frac{e^{-x}}{\sqrt{x}}dx.
\eeq For a power law-type distribution of particles, for example,
$N(\gamma)= N_0\gamma^{-s}$, one can show that for high frequencies
($\nu\gg\nu_c$) the intensity $I_{\nu}$ behaves as \beq\label{inten}
I_{\nu} \propto \int p(\nu,\gamma)N(\gamma)d\gamma\propto
e^{-\nu/\nu_c}, \eeq so that in the case of the Crab the curvature
output might be expected to decay quasi-exponentially for energies
above $\sim 10$ GeV, as indeed suggested by recent very high energy
gamma-ray measurements (\cite{albert}; \cite{aliu}). Inverse Compton
(Klein-Nishina) up-scattering of thermal soft photons, on the other
hand, could result in a (quasi-isotropic) power output of $\sim
10^{31}$ erg/s at TeV energies, consistent with, e.g., existing IACT
constraints on the observed (non-steady) emission from the Crab.

For older pulsars (e.g., $P \sim 1$ s) and $M\gg 1$, co-rotation usually
imposes the strongest constraint, so that achievable maximum Lorentz
factors are typically limited to $\sim (10^4-10^5)$. Although curvature
radiation may then peak in the optical-UV (up to $\sim 10$ eV) and
inverse Compton (Thomson) scattering of curvature or thermal photons
could result in very high energy emission up to $\sim 50$ GeV, their
associated power is negligible due to the small electron Lorentz factors
and the substantially reduced target photon energy density. Hence,
within the approach considered only young pulsars ($P\lppr 0.1$ s)
might be expected to produce detectable high energy gamma-ray
emission.

The proposed scenario could in principle work for a variety of
angles, so that the resultant emission needs not necessarily to be
strongly pulsed. Note that for most circumstances, the major
condition limiting the Lorentz factors of electrons results from
co-rotation. Yet, for large inclination angles, curvature radiation
reaction can become dominant over co-rotation, cf.
Eqs.~(\ref{corotation}) and (\ref{curvgama}) and see Fig.~\ref{fig}
for illustration.

The analysis presented is based on a number of idealizations, which
we plan to remedy in future studies. This particularly involves the
assumptions of, e.g., quasi-straight field lines and a single
particle approach in which plasma effects are neglected.  On the
other hand, one of the strengths of the present concept is its
ability to explicitly take inertial effects into account and so to
allow to estimate the size and extent of the VHE regions in young
pulsars.

\section*{Acknowledgments}
Discussions with Felix Aharonian, George Machabeli and Vasily Beskin
are gratefully acknowledged. Z.O. acknowledges the hospitality of the
Max-Plank Institute for Nuclear Physics (Heidelberg, Germany) during
his short term visits. The study of Z.O. was partially supported by the
Georgian National Science Foundation grant GNSF/ST06/4-096.



\begin{thebibliography}{999}
\bibitem[Aharonian et al. 2007]{ahar} Aharonian F. et al. (HESS Collaboration) 2007, A\&A 466, 543
\bibitem[1990]{abr} Abramowicz M.A. \& Prasanna A.R., MNRAS, 1990, 245, 729
\bibitem[Albert et al. 2008]{albert} Albert J. et al. (MAGIC Collaboration) 2008, ApJ 674, 1037
\bibitem[Aliu et al. 2008]{aliu} Aliu E. et al. (MAGIC Collaboration) 2008, Science 322, 1224
\bibitem[Arons 1983]{arons} Arons J. 1983, ApJ 266, 215
\bibitem[Arons \& Scharlemann 1979]{arons1} Arons J. \& Scharlemann E.T., 1979, ApJ 231, 854
\bibitem[Beskin et al. 1983]{beskin83} Beskin V.S., Gurevich A.V., Istomin Ya.N. 1983, Soviet Phys. JETP 58, 235
\bibitem[Beskin \& Rafikov (2000)]{besk} Beskin V.S. \& Rafikov R.R., 2000, \mnras, 313, 433
\bibitem[Blandford \& Payne 1982]{bp82} Blandford, R. D., \& Payne, D. G., 1982, \mnras, 199, 883
\bibitem[Blumenthal \& Gould (1970)]{blum} Blumenthal G.R., \& Gould R.J., 1970, Rev. Mod. Phys. 42, 237
\bibitem[Cheng et al. 1986]{cheng} Cheng K.S., Ho C., Ruderman M.A., 1986, ApJ, 300, 500
\bibitem[Chedia et al. 1996]{che96} Chedia, O. V., Kahniashvili, T. A., Machabeli, G. Z. \& Nanobashvili, I. S. 1996, Ap\&SS 239, 57
\bibitem[Chiang \& Romani 1994]{chiang} Chiang J. \& Romani R.W. 1994, ApJ 436, 754
\bibitem[Contopoulos et al. 1999]{contop} Contopoulos I., Kazanas D. \& Fendt C.,
 1999, ApJ 511, 351
\bibitem[Crusius-W\"atzel et al. 2001]{crus} Crusius-W\"atzel et al. 2001, ApJ 546, 401
\bibitem[Dalakishvili et al. 2007]{dalak} Dalakishvili G.T., Rogava A.D. \& Berezhiani V.I. 2007, Phys. Rev. D, 76, 045003
\bibitem[Daugherty \& Harding 1982]{dau} Daugherty J.K. \& Harding A.K., 1982, ApJ, 252, 337
\bibitem[Daugherty \& Harding 1983]{dau1} Daugherty J.K., \& Harding A.K. 1983, ApJ 273, 761
\bibitem[Dermer \& Sturner 1994]{der} Dermer C.D. \& Sturner S.J., 1994, ApJ 420, L75
\bibitem[Eikenberry et al. 1997]{eiken} Eikenberry S.S. et al. 1997, ApJ 477, 465
\bibitem[Erber 1966]{erber} Erber T. 1966, Rev. Mod. Phys. 38, 626
\bibitem[1996]{gang96} Gangadhara R.T. 1996, A\&A, 314, 853
\bibitem[Gangadhara \& Lesch 1997]{gl97} Gangadhara R.T. \& Lesch H., 1997, A\&A 323, L45
\bibitem[1968]{gold1} Gold T. 1968, Nature 218, 731
\bibitem[1969]{gold2} Gold T. 1969, Nature 221, 25
\bibitem[Goldreich \& Julian 1969]{gj} Goldreich, P. \& Julian, W.H., 1969, \apj, 157, 869
\bibitem[Hirotani 2007]{hirotani} Hirotani, K. 2007, ApJ 662, 1173
\bibitem[Kuiper et al. 2001]{kuiper} Kuiper L. et al. 2001, A\&A 378, 918
\bibitem[Lessard et al. 2000]{less} Lessard R.W. et al. (Whipple Collaboration) 2000, ApJ 531, 942
\bibitem[Lyne \& Graham-Smith 2006]{lyne} Lyne A.G. \& Graham-Smith F., 2006, Pulsar Astronomy (3rd ed.), Cambridge Univ. Press
\bibitem[1994]{mr94} Machabeli, G.Z., Rogava, A.D. 1994, \pra, 50, 98
\bibitem[Machabeli et al. 2005]{mach05} Machabeli G.Z., Osmanov Z.N., Mahajan, S.M. 2005, Physics of Plasmas 12, 062901
\bibitem[Massaro et al. 2000]{mas00} Massaro E. et al. 2000, A\&A 361, 695
\bibitem[Massaro et al. 2006]{mas06} Massaro E. et al. 2006, A\&A 459, 859\bibitem[Melrose 1998]{mel98} Melrose D.B., 1998, Proc. of APPTC'97 (Toki, Japan), eds. Y. Tomita et al., 96
\bibitem[Michel 1991]{michel} Michel F.C. 1991, Theory of Neutron Star Magnetospheres, Univ. of Chicago Press
\bibitem[Middleditch et al. 1983]{middl} Middleditch J. et al. 1983, ApJ 273, 261
\bibitem[Morini 1981]{mor} Morini M., 1981, Ap\&SS 79, 203
\bibitem[Muslimov \& Tsygan 1992]{musl} Muslimov A.G. \& Tsygan A.I. 1992, MNRAS, 255, 61.
\bibitem[Ochelkov \& Usov 1980]{ochusov} Ochelkov Yu. P. \& Usov V.V. 1980, Ap\&SS 69, 439
\bibitem[Osmanov et al. 2007]{osm7} Osmanov Z., Rogava A.S. \& Bodo G., 2007, \aap, 470, 395
\bibitem[Pacini \& Salvat 1983]{pacini} Pacini, F. \& Salvat M. 1983, ApJ 274, 369
\bibitem[Possenti et al. 2002]{possenti} Possenti A. et al., 2002, A\&A 387, 993
\bibitem[Rieger \& Mannheim 2000]{rm00} Rieger, F. M., \& Mannheim, K. 2000, \aap, 353, 473
\bibitem[Rogava et al. 2003]{r03} Rogava A. D., Dalakishvili G. \& Osmanov Z., 2003, Gen. Rel. and Grav. 35, 1133
\bibitem[Ruderman \& Sutherland 1975]{rud} Ruderman A. \& Sutherland P.G., 1975, ApJ., 196, 51
\bibitem[Ruderman 1972]{rud72} Ruderman, M. 1972, ARA\&A 10, 427
\bibitem[Shearer \& Golden 2001]{shear} Shearer A. \& Golden A., 2001, ApJ 547, 967
\bibitem[Sollerman 2003]{soller} Sollerman J., 2003, A\&A 406, 639
\bibitem[Tsai \& Erber 1974]{tsai} Tsai W-Y. \& Erber T., 1974, Phys. Rev. D 10, 492
\bibitem[Thomas \& Gangadhara 2007]{thom07} Thomas, R.M.C. \& Gangadhara R.T. 2007, A\&A 467, 911
\bibitem[Tsuruta et al. 2002]{tsu} Tsuruta S. et al. 2002, ApJ, 571, L143
\bibitem[Usov \& Shabad 1985]{usov} Usov V.V. \& Shabad A., 1985, Ap\&SS 117, 309
\bibitem[Yakovlev \& Pethick 2004]{yakpet} Yakovlev D.G. \& Pethick C.J. 2004, ARA\&A 42, 169
\bibitem[Zhang \& Harding 2000]{zhang} Zhang B. \& Harding A.K., 2000, ApJ, 532, 1150


\end{thebibliography}
\end{document}